\documentclass[runningheads]{svmult}

\usepackage{makeidx}   
\usepackage{graphicx}  
\usepackage{subeqnar}  
\usepackage{multicol}  
\usepackage{physprbb}  
\makeindex             

\def\lsim{\raise0.3ex\hbox{$<$}\kern-0.75em{\lower0.65ex\hbox{$\sim$}}}
\def\gsim{\raise0.3ex\hbox{$>$}\kern-0.75em{\lower0.65ex\hbox{$\sim$}}}

\begin{document}
\title*{Theoretical Implications of the Gamma-Ray Burst -- Supernova Connection}
\titlerunning{Theoretical Implications of the GRB--SN Connection}
%
\author{Roger~A.~Chevalier}
\authorrunning{}
%
%
\institute{Department of Astronomy, University of Virginia,\\
Charlottesville, VA 22903, USA}

\maketitle              

\begin{abstract}
Although observations of circumstellar shock interactions around supernovae are
generally consistent  with a $\rho\propto r^{-2}$ wind surrounding
  the progenitor star, this is not  true for GRB (gamma-ray burst)
afterglows.
However, GRB 991208 and GRB 000301C may be consistent with wind interaction
if the injection particle spectrum is a broken power law.
Circumstellar dust echos can place  constraints on 
supernova and GRB progenitors, but have been clearly observed only
around SN 1987A.
Excess emission observed in two GRB afterglows is more likely to have
 a supernova origin.
An interstellar dust echo, causing the light curve to flatten out, is
a possibility for GRB afterglows, but is not likely to be observable.
\end{abstract}

\section{Introduction}
A year ago, there were a number of pieces of evidence that pointed
to a connection between GRBs (gamma-ray bursts) and SNae (supernovae)
(see \cite{K00} for a review).
The positions of GRBs in galaxies, near star forming regions, were
consistent with a massive star origin.
In addition to the association of GRB 980425 with SN 1998bw, two
distant GRB afterglows showed evidence for supernovae in their
light curves and spectra.
In some cases, the afterglow evolution suggested interaction with
a freely expanding stellar wind, as expected in the immediate vicinity
of a massive star.

Although the evidence for an association with star forming regions
has held up \cite{Bl00}, the other topics have not provided further
support.
There have been no further convincing associations of supernovae
with GRB afterglows.  Dust echos have been proposed as the
mechanism for the two apparent distant supernova cases that
have been observed \cite{WD00,EB00}.
The two best  observed recent afterglows, GRB 991208 and GRB 000301C,
have been interpreted in terms of expansion in a constant density
medium, with jet effects \cite{G00,B00}.
All these points weaken the case for a GRB -- SN connection.
Here, I discuss these recent developments in the context
of studies of supernovae and GRB afterglows.

\section{Supernova and GRB Shock Interactions}

The shock interactions of supernovae with circumstellar matter
 can be observed through radio synchrotron
emission (analogous to the synchrotron emission in GRB afterglows)
and through X-ray thermal radiation.
The radio light curves are characterized by an initial rise followed
by a power law decline.
A plot of the time of the light curve peak vs. the peak luminosity
at that time sorts the observed supernovae into various types \cite{C98,LC99}.
These properties are related to the density of the gas into which the
shock wave is propagating.
Except for SN 1987A,  Type II supernovae are thought
 to have red supergiant progenitors, which have
wind velocities of $\sim 10$ km s$^{-1}$.
The inferred mass
loss rate from the progenitor of SN 1999em, a recently detected radio
supernova \cite{L99},
 is $\dot M \sim 10^{-6}~M_\odot~\rm yr^{-1}$.
This may be typical of SN II progenitors; the well-observed Type II 
radio supernovae are probably 
  at the upper end of a distribution of circumstellar
densities.
The most luminous radio supernovae are inferred to have 
$\dot M\sim   10^{-4}-10^{-3}~M_\odot~\rm yr^{-1}$.

Although some deviation from standard evolution
 has been found in radio supernovae
 with red supergiant progenitors \cite{M98,M00}
the variations are small compared to those in SN 1987A.
The well-known equatorial ring is at a radius of $6.3\times 10^{17}$ cm,
but moderately dense gas probably extends in from this radius because
of photoionization by the progenitor star \cite{CD95}.
The radio luminosity was initially low compared to the red
supergiant case because of interaction
with the low density progenitor wind, but started to increase in 1990
and rapidly rose to $> 100$ times an extrapolation of the early
decline \cite{B95}.
During the last few years, the supernova shock wave has started to
interact with parts of the equatorial ring that are closest to the
supernova \cite{LS00}.
The changes in radio flux, which are more dramatic than any observed for
other Type II supernovae, show what happens when a supernova shock wave
runs into dense gas from an earlier evolutionary stage.
At the present time, SN 1987A remains unique among the SN IIae and
the precise causes for the progenitor evolution to the blue are not clear;
  they may have to do with binary evolution.

Except for SN 1998bw, the SN Ib/c that have been observed in the radio
have similar properties \cite{LC99,K98}.
If they are able to efficiently radiate synchrotron emission, the
radio flux is approximately in accord with
$\dot M\sim  10^{-5}-10^{-4}~M_\odot~\rm yr^{-1}$ and
$v_w \approx 1,000$  km s$^{-1}$, as expected for a Wolf-Rayet star
\cite{C98}.
Except for SN 1998bw, the radio light curves show a smooth evolution,
although the data are sparse.
In the case of SN 1998bw, the observed rise in flux between days 20 and
30 \cite{K98} is more likely to be due to an increase in the energy
of the explosion than to an encounter with denser gas \cite{LC99}.
The evolution is consistent with interaction with a
$\rho\propto r^{-2}$ wind extending out to at least $4\times 10^{17}$ cm.

If GRBs have massive star progenitors, they are likely to be
Wolf-Rayet stars.
Arguments in favor of this assumption include: (1)  SN 1998bw, the best
case of a SN -- GRB association (GRB 980425), was of Type Ic, with a
probable Wolf-Rayet progenitor.
(2)  The high energy of GRBs suggest that a moderately massive 
black hole is involved, which, in turn, requires a massive, $\gsim
 20-25 ~M_\odot$,
progenitor \cite{EH98}.
These stars are likely to be Wolf-Rayet stars at the end of their 
lives \cite{GS96}.
(3)  The relativistic flow from a central object may be able to penetrate
a relatively compact Wolf-Rayet star, but probably cannot
penetrate an extended red supergiant star \cite{MW99}.
None of these arguments is definitive, but they are suggestive.
The GRB rate is a small fraction of the SN rate so
that a peculiar kind of star with an unusual surroundings could be
involved.
For example, a binary merger may be needed so that the central black hole
has a large rotational energy.
The merger process may be associated with a strong equatorial outflow
of the stellar envelope.
However, the relativistic flow is likely to be along the polar axis,
where the mass loss properties may be normal.

The best tests for afterglow models are sources with extensive data;
radio data are especially useful because they give information on
the absorption frequency and the peak flux and its frequency.
The extensive radio data \cite{FWK00}
on GRB 970508 can be fitted by a model afterglow interacting with
an $r^{-2}$ wind, extending out to $3\times 10^{18}$ cm \cite{CL00}.
Frail et al. \cite{FWK00} fit the same data by jet expansion in
a uniform medium.
In their model, jet effects become important at day 25 and there is a
transition to spherical nonrelativistic flow at day 100.
The radio data on GRBs 991208 and 000301C can also be approximately
fitted by a wind model \cite{LC01}, but in these cases a broken power
law spectrum is needed for the particles in order to fit the
higher energy light curves.
The jet in a uniform medium models for these objects can assume a
single power law spectrum; however, the evolution of  GRB 991208 is
taken to be in the jet transition phase during the period of
observation ($\sim 3-20$ days) \cite{G00}, whereas  GRB 000301C  is
taken to have a sharp transition to asymptotic jet evolution on day 7.3
\cite{B00}.
A problem with the jet models is that the evolution is not well defined.
The broken power law, wind models can be tested by late observations
because different time dependences are expected at radio and optical
wavelengths.
The wind interaction models \cite{CL00,LC01} have not included
jet effects, although a collimated flow is expected if it must
escape from the center of a star.
To some extent, this can be justified by the slow apparent evolution of
a jet in a wind \cite{KP00,GD00}.

The facts that many afterglows can be fitted by model evolution in a
constant density medium and that massive stars are attractive progenitors
have led to the suggestion \cite{W01} that the expansion is into the
constant density medium that is expected downstream from the termination
shock of the massive star wind \cite{CL99}.
The outer radius of such a region
is $\sim 2-2.5$ times the inner radius  when
the fast wind from a Wolf-Rayet star runs into a slow wind from
a previous evolutionary stage \cite{CL99,CI83}.
Models and observations of Galactic Wolf-Rayet stars show that the
swept-up shell of red supergiant material at the outer radius is
at a distance $\gsim 3$ pc from the star \cite{GS96,CL99}.
This radius is sufficiently large that interaction with
the free $\rho\propto r^{-2}$ wind is expected over the typical period of 
observation of afterglows.
The time for the wind to reach the termination shock is relatively
short ($\sim 300$ years to reach $10^{18}$ cm), so that the assumption
of constant wind properties is plausible.
In a massive star progenitor model, a high density in the immediate
vicinity of the explosion is expected even if the wind passes through
a termination shock at a relatively small radius.
The prompt optical emission from a GRB gives the opportunity to 
investigate the immediate surroundings; in the case of GRB 990123,
low density (interstellar medium) interaction gives a better
explanation for the observations \cite{CL00,SP99}.

Despite the plausibility of free wind interaction, the uncertainty
in the evolution of massive stars leaves open the possibility
of interaction with denser material at early times; 
Ramirez-Ruiz et al. \cite{RR00} have investigated such a case.
The interaction of a GRB flow with a dense shell could have different
properties compared to a supernova, as exemplified by SN 1987A.
In the supernova case, the shock front is 
continually driven by the lower velocity
supernova ejecta with most of the kinetic energy.
Most GRB afterglow models have constant energy, so that although
an initial overpressure can be expected upon a collision with a
shell, it is not as marked as in the supernova case.
In addition, the GRB flow is likely to be in a jet, unlike the supernova
case.
When the beamed flow interacts with a large density jump, it is rapidly
decelerated and lateral spreading can occur.
The afterglow could make a rapid transition to the asymptotic
behavior for a jet flow \cite{SPH99}.
The afterglow of GRB 000301C did show a bump followed by a transition
to a steeper decline that could be lateral jet expansion, although
the bump can be interpreted as a microlensing event \cite{GLS00}.

\section{Dust Echos}

The possibility of circumstellar dust echos arises naturally for
Type II supernovae with red supergiant progenitors because their
cool winds are known to contain dust.
For the higher estimated mass loss rates, e.g., $\sim
10^{-4}~M_\odot~\rm yr^{-1}$ for SN 1979C, the progenitor star
is expected to be entirely enshrouded in dust.
Yet when we observe Type II supernovae, there is little evidence
for dust absorption in the spectra of the supernovae, including SN 1979C.
The probable solution is that the dust near the star is evaporated
by the radiation from the supernova and the remaining dust does not
have a large optical depth.
The radius at which dust becomes so hot that it is evaporated, $r_v$,
occurs at $\sim 3\times 10^{17}$ cm for typical supernova and
dust parameters \cite{D83}.

The circumstellar dust beyond $r_v$ can give rise to infrared and
scattered light echos.  
Both SN 1979C and SN 1980K showed evidence for infrared excesses that
might be attributable to dust echos \cite{D83}.
The late optical light from these supernovae appeared to be dominated by
emission from the circumstellar shock wave interactions and did
not show dust scattering effects \cite{C86}.
Roscherr and Schaefer \cite{RS00} have recently examined the emission from 
SN IIn to search for scattered light echos, but again found
that the late emission is dominated by circumstellar shock interactions.
Circumstellar scattered light echos have been observed around SN 1987A
\cite{Cr95} because it was possible to spatially resolve the echos
away from the supernova; it would not have been possible to observe
the echos for a supernova at a typical distance.

No evidence for circumstellar dust echos has been found in Type Ib/c
supernovae, which is not surprising considering their presumed
Wolf-Rayet star progenitors.
Wolf-Rayet stars atmospheres are too hot for dust formation.
Despite this, circumstellar dust has been found in a small
fraction of these stars.
The source appears to be the compressed interaction region where the Wolf-Rayet
star wind collides with the wind from a close companion,
as indicated by the pinwheel dust patterns observed around 
two Wolf-Rayet stars \cite{MTD99}.
Dust may be more commonly present in circumstellar 
shells at radii $\gsim 3$ pc, but such shells would have only 
a small dust optical depth.

Circumstellar dust echos have been of recent interest for GRB afterglows
because of the possibility that they could explain the excess emission
observed in two sources at ages $10-60$ days without the need for
a supernova \cite{WD00,EB00}.
An echo from thermal re-radiation by dust is too red to explain the
observations, so the scattered light explanation is considered here \cite{EB00}.
The timescale for the emission sets the dust at a radial distance of
$4.5\times 10^{17}$ cm and its color requires an optical depth at
3000 \AA\ of 7  
\cite{EB00}.
The required dust mass is $\sim 0.1 ~M_\odot$ or $\sim 10 ~M_\odot$ 
total mass in the
shell.
If the gas is moving out as part of the stellar wind, special timing
is needed to have much of the envelope mass end up at the appropriate
radial distance.
Thus, on the basis of the fact that the most plausible massive star 
progenitors for GRBs (Wolf-Rayet stars) do not typically have dusty winds 
 plus the special conditions that are needed for the dust if present,
the supernova hypothesis appears to be a more natural explanation
for the  excess light than a dust echo.
Reichart \cite{R00} has recently examined detailed models for circumstellar
echos and found that the colors of the excess light of GRB 970228
are inconsistent with an echo.
Even if the excess light could be explained as a scattered light echo,
the implication would be that the progenitor object was a massive star
which was a red supergiant  at the time of the explosion or soon before
the explosion.

In addition to circumstellar dust echos, dust in interstellar clouds
can give rise to echo phenomena.
In the supernova case, the well-known rings around SN 1987A are
due to clouds that are 100 pc or more in front of the supernova
\cite{CE88}.
Similar scattered light echos have been observed around the
Type Ia supernovae SN 1991T and SN 1998bu  \cite{Sp99,Ca00}.
The echo phenomenon is characterized by a flattening of the supernova
light curve, a late spectrum related to the spectrum near maximum light,
and a significant spatial size related to the distance of the interstellar
cloud in front of the supernova.
If the peak luminosity, $L_p$, of the supernova or GRB lasts for a time
$\Delta t$, the ratio of the echo luminosity to $L_p$ is
$\tau c F\Delta t /2d$, where $\tau$ is the dust optical depth, $c$ is
the speed of light, $F$ is the phase function (forward scattering
is expected), $d$ is the distance of the cloud in front of the
explosion, and single scattering is assumed \cite{CE88}.
This can explain the $9-10$ magnitude decline from peak to flattening
of the light curve in the 3 observed echos.
GRB 990123 has the best defined afterglow optical light curve; it
peaked at $\sim 9$ mag. and subsequently declined more rapidly than
$t^{-1}$ so that most of the radiated energy appeared near the peak.
If a similar cloud were placed in front of GRB 990123 to those in front
of SNae 1991T and 1998bu, the fact that $\Delta t\sim 30$ sec for
GRB 990123, but is $\sim 20$ days for the supernovae would make
the luminosity $\sim 6\times 10^4$ fainter compared to the peak luminosity.
This would give an echo magnitude $\sim 30$, which is undetectable.
The angular radius of the echo from the explosion is $0.08 (t/{\rm yr})^{1/2}
(d/{\rm 100~pc})^{-1/2}$ rad, which could be outside the beamed emission
from a GRB, further reducing the echo brightness.
It is only for a relatively nearby GRB with a favorable distribution
of dust that there is any chance of detecting an echo.
The best candidates for an echo would be sources that show evidence
for dust in the host galaxy, e.g., GRB 000418 \cite{Kl00}.
Such a detection would be of interest in that it would give information on the
very early optical luminosity of the burst.

\medskip
I am grateful to C. Fransson and Z.-Y. Li for discussions and
collaboration, and to NASA grant NAG5-8130 for support.

%

\end{document}